\documentclass[10pt,twocolumn]{revtex4}
\usepackage{amssymb,epsf}
\usepackage{latexsym}                                  
\usepackage{graphicx}
\usepackage{float}
\begin{document}
\title{Thermodynamic behavior and stability of Polytropic gas}
\author{H. Moradpour$^a$\footnote{h.moradpour@riaam.ac.ir}, A. Abri$^a$,
H. Ebadi$^{a,b}$\footnote{hosseinebadi@tabrizu.ac.ir}}
\address{$^a$Research Institute for Astronomy and Astrophysics of Maragha (RIAAM),
P.O. Box 55134-441, Maragha, Iran,\\
$^b$Astrophysics Department, Physics Faculty, University of
Tabriz, Tabriz, Iran.}

\begin{abstract}
We focus on the thermodynamic behavior of Polytropic gas as a
candidate for dark energy. We use the general arguments of
thermodynamics to investigate its properties and behavior. We find
that a Polytropic gas may exhibit the dark energy like behavior in
the large volume and low temperature limits. It also may be used
to simulate a fluid with zero pressure at the small volume and
high temperature limits. Briefly, our study shows that this gas
may be used to describe the universe expansion history from the
matter dominated era to the current accelerating era. By applying
some initial condition to the system, we can establish a relation
between the Polytropic gas parameters and initial conditions.
Relationships with related works has also been addressed.
\end{abstract}

\maketitle
\section{Introduction}
In the Einstein General Relativity framework,
Friedmann-Robertson-Walker metric (FRW) and its conformal form
help us get a suitable model for describing the universe expansion
and its inhomogeneities in scales smaller than $100-$Mpc,
respectively \cite{roos,rma}. In standard cosmology, the universe
is born from a singularity called big bang and is inflated in
primary moments. Thereinafter, the rate of this expansion is
decreased, since Radiation and Matter play the role of dominated
fluid, which is the determinant of the universe expansion rate, in
subsequent eras \cite{roos}. Finally, the universe deals with an
accelerated expansion with positive rate, about $13700-$Myr after
big bang \cite{roos,Rie,Rie1,Rie2,Rie3}. Although the latest and
current phase of universe expansion satisfies the thermodynamics
stability conditions \cite{noj,noj1,pavon2,msrw,mr} but, the
nature of dominated fluid, which supports this phase of expansion
and called dark energy (DE), is a mysterious puzzle
\cite{rev1,rev2,rev3}.

In addition to the mentioned puzzle, standard cosmology suffers
from a series of other weaknesses such as the coincidence and fine
tuning problems \cite{roos}. There are various attempts to give a
candidate for describing DE and solving the standard cosmology
problems \cite{rev1,rev2,rev3}. Among these attempts, some models
include a varying DE candidate with equation of state
$P=\rho+f(\rho)$, where $P$ and $\rho$ are the DE pressure and
energy density, respectively \cite{noj2,noj3}. In astrophysics, a
Polytropic gas with equation of state
\begin{equation}\label{EoS}
P=K \rho ^{(1+\frac{1}{n})},
\end{equation}
where
\begin{equation}\label{energy density}
\rho =\frac{U}{V},
\end{equation}
is its energy density, while $U$ and $V$ are the total energy of
gas and the volume of its container, respectively, has numerous
applications \cite{cris}. In addition, $n$ called the Polytropic
index and $K$ is a constant \cite{cris}. An adiabatic gas
\cite{CALEN} and a degenerate electron gas \cite{cris} are two
examples of such gases. Since the equation of state of this gas is
in the $P=f(\rho)$ form, it attracts the attention of cosmologist
to itself as a probable way to describe DE and the coincidence as
well as the fine tuning problems
\cite{karami1,karami2,karami3,karami4,ahep,ansari,ref56,ref57,ref58,ref59,salti,taji,sarkar,in,gref1,nK}.

Karami et al. investigated the mutual interaction between
Polytropic gas (as the DE candidate) and CDM, and find out that
for some even and positive Polytropic index and some positive
values of $K$, the Polytropic gas behaves as the Phantom dark
energy \cite{ref59}. Moreover, the generalized second law of
thermodynamics is always satisfied by a universe filled with a
Polytropic gas model of DE together with a CDM which interact with
each other \cite{karami4}. Authors in \cite{in} used the first law
of thermodynamics, and qualitatively show that a Polytropic gas
may exhibit a DE behavior. In addition, they show that this model
leads to a suitable fitting with observational data about the
current expanding era \cite{in}. In fact, Asadzadeh et al. have
argued that a Polytropic gas model of DE with Polytropic index
$n<-1$ and $K<0$ is in suitable agreement with the observational
data \cite{nK}. It is also shown that the Polytropic gas model of
DE may simulate the new agegraphic as well as the holographic dark
energy models and thus the current accelerating phase of universe
\cite{karami3,taji}. Interaction between the Polytropic gas model
of DE and dark matter in Kaluza-Klein cosmology has also been
studied \cite{gref1}.

It is useful to mention here that in all of Refs.
\cite{karami1,karami2,karami3,karami4,ahep,ansari,ref56,ref57,ref58,ref59,salti,taji,sarkar,in,nK,gref1},
authors investigate the Polytropic gas properties and behaviors by
considering the cosmological setup. As a common result, all of
them find that, in the FRW background, the Polytropic gas may be a
suitable candidate for describing DE. Therefore, one may ask that
what is the origin of the DE like behavior of Polytropic gas? Does
this ability of the Polytropic gas to describe DE is due to the
gravitational theory used to describe the universe expansion?
Indeed, does thermodynamical properties of Polytropic gas is
similar with those of DE without considering a cosmological
background? Such confusions have been arisen in studying the
Chaplygin gas as well as its generalized and modified forms
\cite{plb1,plb2}. In fact, authors in \cite{plb1,plb2} by focusing
on the thermodynamic arguments, have been shown that such gases
have enough ability for exhibiting the DE like behavior. Although,
the equation of state of Polytropic gas is similar to that of the
generalized Chaplygin gas but the physics behind them are
completely different. Whiles, the Chaplygin gas model comes from
the Russian physicist attempts to study the lifting force which
applies on the plane wings in aerodynamics \cite{chap}, the
Polytropic gas models are introduced to study the hydrostatic
equilibrium equation and thus the star evolution \cite{cris}.
Therefore, it is not straightforward to generalize the results of
Refs.~\cite{plb1,plb2} to Polytropic gas models.

Here, we want to study the thermodynamics of a Polytropic gas
confined to volume $V$. In fact, we are eager to know that under
which thermodynamic conditions a Polytropic gas exhibits the DE
like behavior. Does this model satisfies the thermodynamic
stability conditions, and does it pleases the thermodynamic
expectations? Our study shows that the Polytropic gas may
naturally exhibit the DE like behavior. Since we only use
thermodynamics to study the behavior of Polytropic gas, our
results are independent of considering the gravitational theory
used to describe the universe expansion.

The paper is organized as follows. In the next section, we
consider an adiabatic Polytropic gas and evaluate its
thermodynamical properties such as its energy as a function of
volume. The thermodynamic stability conditions are also addressed.
Section $III$ includes more debates about the thermodynamic
stability conditions and some examples. In section $IV$, by
applying initial conditions to the system we get a relation
between the Polytropic index, $K$ and initial conditions. Section
$V$ includes summary and concluding remarks.
\section{Adiabatic Polytropic gas EoS}
The pressure of a fluid with energy $U$ confined into a cylinder
with volume $V$ is evaluated as \cite{CALEN}
\begin{equation}\label{thermodynamics}
\left(\frac{\partial U}{\partial V}\right)_{S}=-P.
\end{equation}
Combining Eqs.~(\ref{EoS}) and~(\ref{energy density}), and
inserting the result into~(\ref{thermodynamics}) to get
\begin{equation}\label{thermodynamics1}
\left(\frac{\partial U}{\partial
V}\right)_{S}=-K\left(\frac{U}{V}\right)^{1+\frac{1}{n}}.
\end{equation}
By taking integral from this equation, one finds the energy of
Polytropic gas as
\begin{equation}\label{thermodynamics2}
U=(-1)^{-n}\left(KV^{-\frac{1}{n}}+b\right)^{-n},
\end{equation}
where $b=b(S)$. It is also apparent that
Eqs.~(\ref{thermodynamics1}) and~(\ref{thermodynamics2}) are also
available, if $K=K(S)$. By defining $\delta
^{-\frac{1}{n}}=\frac{b}{K}$ and simple calculations, we get
\begin{equation}\label{thermodynamics3}
U=(-1)^{-n}K^{-n}V(1+(\frac{\delta}{V})^{-\frac{1}{n}})^{-n},
\end{equation}
which leads to
\begin{equation}\label{8}
\rho=\frac{U}{V}=(-1)^{n}K^{-n}(1+(\frac{V}{\delta})^{\frac{1}{n}})^{-n},
\end{equation}
for the energy density of Polytropic gas. This equation leads
to
\begin{equation}\label{80}
\rho\sim(-1)^{n}K^{-n},
\end{equation}
and
\begin{equation}\label{81}
\rho\sim(-1)^{n}K^{-n}\frac{\delta}{V},
\end{equation}
for $n>0$ and $n<0$, respectively, in the small volume limits.
Since it seems that density should be positive for even $n$
\cite{ref59,taji,cris}, these equations state that $\delta$ should
meet the $\delta>0$ condition. Since $\delta>0$, one can use
Eq.~(\ref{8}) to conclude that, independent of $n$, density is
positive for $K<0$. Moreover, inserting this equation
into~(\ref{EoS}) to reach
\begin{equation}\label{11}
P=(-1)^{n+1}K^{-n}(1+(\frac{V}{\delta})^{\frac{1}{n}})^{-(n+1)},
\end{equation}
which yields
\begin{eqnarray}\label{press}
P=\frac{-\rho}{(1+(\frac{V}{\delta})^{\frac{1}{n}})},
\end{eqnarray}
and
\begin{eqnarray}\label{press0}
\omega=\frac{P}{\rho}=-\frac{1}{(1+(\frac{V}{\delta})^{\frac{1}{n}})},
\end{eqnarray}
as the corresponding pressure and the state parameter of
Polytropic gas, respectively. Therefore, the state parameter of
Polytropic gas meets the $-1\leq \omega \leq 0$ condition. It is
interesting to note that for $b=0$, one obtains
\begin{eqnarray}\label{press1}
P=-\rho=(-1)^{n+1}K^{-n},
\end{eqnarray}
and
\begin{eqnarray}\label{press2}
\omega=\frac{P}{\rho}=-1,
\end{eqnarray}
which are independent of the system volume. Therefore, the $b=0$
case is capable to explain the primary inflationary era and the
current expanding phase as well as the anti de-Sitter spacetime.
For example, consider a situation in which $K>0$. While
$\frac{dK}{dS}=0$, on one hand, for even $n$ this equation is
compatible with the equation of state of the cosmological constant
($\rho>0$) which may support the de-Sitter spacetime. On the other
hand, an odd $n$ points to a fluid with $\rho<0$ which may support
the anti de-Sitter spacetime.

Whenever $n>0$ and $\delta$ does not diverge, one gets
\begin{eqnarray}\label{smv}
P \approx (-1)^{n+1}K^{-n}\simeq-\rho,
\end{eqnarray}
for small volumes, which is the same as the result of the $b=0$
case~(\ref{press1}). In order to derive this equation, we used
Eqs.~(\ref{8}) and~(\ref{press}). This result is in accordance
with the primary inflationary era. It is interesting to note that
the generalized Chaplygin gas has similar behavior at large
volumes \cite{plb1}. For large volumes, Eqs.~(\ref{8})
and~(\ref{press}) lead to
\begin{eqnarray}
\rho \approx \frac{(-1)^{n}K^{-n}\delta}{V}
\end{eqnarray}
and
\begin{eqnarray}
P\approx
\frac{(-1)^{n+1}K^{-n}\delta^{\frac{n+1}{n}}}{V^{\frac{n+1}{n}}},
\end{eqnarray}
respectively. We should note that since $V\gg1$ and $n>0$,
$\omega\sim0$. In fact, while length is proportional to the
cosmological scale factor $a$, $V\propto a^3$ and we get
\begin{eqnarray}\label{smv4}
\rho \approx \frac{(-1)^{n}K^{-n}}{a^3},\ \ P \sim 0,
\end{eqnarray}
which indicates a pressureless matter. This behavior of Polytropic
gas at large volume is the same as that of the generalized
Chaplygin gas in small volumes \cite{plb1}. Therefore, a
Polytropic gas with $n>0$ and $b\neq0$ cannot simulate a fluid
with $P=-\rho\neq0$ at large volumes. Briefly, a Polytropic gas
with $n>0$, while $\delta$ does not diverge, behaves as a fluid
with $\omega\simeq-1$ and a pressureless matter at the small and
large volumes limits, respectively.

Meanwhile, while $n<0$ and $\delta$ does not diverge, by using
Eqs.~(\ref{8}) and~(\ref{press}) one gets
\begin{eqnarray}\label{smv1}
P\sim0,\ \ \rho \approx \frac{(-1)^{n}K^{-n}}{a^3},
\end{eqnarray}
and
\begin{eqnarray}\label{smv2}
P \approx (-1)^{n+1}K^{-n}\simeq-\rho,
\end{eqnarray}
as the pressure and density of Polytropic gas in small and large
volumes, respectively. In order to get~(\ref{smv1}), we assumed
again $V\propto a^3$. Therefore, a Polytropic gas with $n<0$
behaves as the pressureless matter and a fluid with
$\omega\simeq-1$ at small and large volumes limits, respectively.
This behavior is fully consistent with that of the generalized
Chaplygin gas \cite{plb1}. Let us summarize the above results. The
$b=0$ case indicates a fluid with $\omega=-1$ and therefore, may
be used to simulate the primary and current inflationary eras.
Moreover, since the state parameter of a Polytropic gas with $n>0$
is increased from $\omega\simeq -1$ (for small volumes) to
$\omega\simeq 0$ (for large volumes), it may be useful to describe
an expanding universe which expands from an inflationary like
primary era to a matter dominated like era. Finally, since the
state parameter of a Polytropic gas with $n<0$ is decreased from
$\omega\simeq 0$ (for small volumes) to $\omega\simeq -1$ (for
large volumes), it may be useful to describe the universe
expansion from the matter dominated era to the current expanding
phase.

The thermodynamic stability condition or the convexity of the
energy surface requires that \cite{CALEN}
\begin{equation}\label{18}
(\frac{\partial P}{\partial V})_{S} \leq 0,
\end{equation}
and
\begin{equation}\label{188}
C_P\geq C_V\geq0,
\end{equation}
where $C_V$ and $C_P$ are the heat capacities at constant volume
and pressure, respectively. Additionally, since for a fluid with
$N$ particle
\begin{equation}\label{1888}
C_P-C_V=\frac{TV\alpha^2}{N\kappa_T},
\end{equation}
where $\kappa_T=-\frac{1}{V(\frac{\partial P}{\partial V})_{T}}$
and $\alpha$ are the isothermal compressibility and the
coefficient of thermal expansion, respectively \cite{CALEN}, if
$\kappa_T>0$ and $C_V\geq0$ are simultaneously satisfied
condition~(\ref{188}) is also met. The latter means a
thermodynamic system, which satisfies Eq.~(\ref{18}),
\begin{equation}\label{20}
C_{V}>0,
\end{equation}
and
\begin{equation}\label{21}
(\frac{\partial P}{\partial V})_{T} \leq 0,
\end{equation}
also meets the thermodynamic stability condition. For the $b=0$
case, from Eq.~(\ref{press1}), simple calculations lead to
$(\frac{\partial P}{\partial V})_{S}=0$ and therefore, the
stability condition~(\ref{18}) is marginally satisfied. Now, using
Eq.~(\ref{11}) to reach
\begin{equation}\label{19}
(\frac{\partial P}{\partial V})_{S}
=-\frac{\delta^{-\frac{1}{n}}(1+\frac{1}{n})\frac{P}{V}}{1+(\delta
V)^{-\frac{1}{n}}}.
\end{equation}
Bearing the $\delta>0$ condition together with Eqs.~(\ref{8})
and~(\ref{press}) in mind, a Polytropic gas with $\rho>0$
satisfies condition~(\ref{18}), if $-1\leq n<0$. In addition, a
Polytropic gas with negative density ($\rho<0$) satisfies
condition~(\ref{18}) if its Polytropic index meets either the
$n\leq-1$ or $n>0$ conditions. It is also apparent that, just the
same as the $b=0$ case, the $n=-1$ case marginally satisfies
condition~(\ref{18}). In the next section, we investigate the
quality of validity of Eqs.~(\ref{20}) and~(\ref{21}).
\section{Thermal Polytropic gas EoS}
One should determine the thermal equation of state $P=P(T,V)$ to
investigate the quality of validity of Eq.~(\ref{21}). In
addition, since
\begin{equation}\label{37}
C_{V}=T(\frac{\partial S}{\partial T})_{V},
\end{equation}
we need also to determine the thermal equation of state $S=S(T,V)$
in order to study the behavior of $C_V$ \cite{CALEN}. Here, since
in cosmological setups authors generally set $K$ to a constant
value \cite{cris,ref59,taji}, we only consider the situation in
which $\frac{dK}{dS}=0$ which means that $K$ is constant. In order
to evaluate the temperature of Polytropic gas, inserting
Eq.~(\ref{thermodynamics3}) into
\begin{equation}\label{23}
T=(\frac{\partial U}{\partial S})_{V},
\end{equation}
and get
\begin{equation}\label{24}
T=(-1)^{n+1}nV^{1+\frac{1}{n}}(K+bV^{\frac{1}{n}})^{-(n+1)}\frac{db}{dS},
\end{equation}
as the temperature of Polytropic gas. Bearing Eqs.~(\ref{8})
and~(\ref{press}) together with the definition of $\delta$ in
mind, this equation can be written as
\begin{equation}\label{244}
T=-n\frac{\rho
V^{1+\frac{1}{n}}}{(1+(\frac{V}{\delta})^{\frac{1}{n}})}\frac{db}{dS}.
\end{equation}
In order to continue our analysis, we need an exact form for $b$.
To achieve this aim, bearing Eq.~(\ref{thermodynamics2}) in mind,
a dimensional analysis shows
\begin{equation}\label{31}
[b]^{-n}=[U].
\end{equation}
In addition, from Eq.~(\ref{244}) we have $db\propto dS$. Finally,
since $[U]=[TS]$ in thermodynamics \cite{CALEN}, a primary simple
election which satisfies Eq.~(\ref{31}) is
\begin{equation}\label{32}
b=(T_{*} S)^{-\frac{1}{n}},
\end{equation}
where $T_*$ is a universal constant with temperature dimension,
which should be evaluated from other parts of physics such as
statistical mechanics or experimental data. Such analysis yields
similar results for generalized Chaplygin gas and can be found in
Ref. \cite{plb1}. Taking derivative with respect to $S$ from this
equation to get
\begin{equation}\label{33}
\frac{db}{dS}=-\frac{1}{n}T_{*}^{-\frac{1}{n}}S^{-\frac{1}{n}-1}.
\end{equation}
Inserting this result into~(\ref{244}) to obtain
\begin{equation}\label{2444}
T=\frac{\rho
V^{1+\frac{1}{n}}}{(1+(\frac{V}{\delta})^{\frac{1}{n}})}T_{*}^{-\frac{1}{n}}S^{-\frac{1}{n}-1}.
\end{equation}
Now, inserting Eqs.~(\ref{8}) and~(\ref{32}) into~(\ref{2444}) to
get
\begin{equation}\label{34}
T=(-1)^{n}V^{1+\frac{1}{n}}(T_{*}^{-\frac{1}{n}}S^{-\frac{1}{n}-1})[K+T_{*}^{-\frac{1}{n}}S^{-\frac{1}{n}}V^{\frac{1}{n}}]^{-(n+1)},
\end{equation}
which leads to
\begin{equation}\label{35}
S=[(\frac{T_{*}}{T})^{\frac{1}{n+1}}(-1)^{\frac{n}{n+1}}-1]^{n}\frac{V}{K^{n}T_{*}},
\end{equation}
for the entropy of Polytropic gas. By combining
Eqs.~(\ref{press}),~(\ref{2444}) and~(\ref{35}) one gets
\begin{equation}\label{39}
P=K^{-n}(-1)^{n+1}(1+[(\frac{T_{*}}{T})^{\frac{1}{n+1}}(-1)^{\frac{n}{n+1}}-1]^{-1})^{-(n+1)},
\end{equation}
as the last thermal equation of state ($P=P(T,V)$) needed to
investigate the system. From this equation it is apparent that
$(\frac{\partial P}{\partial V})_T=0$ which indicates
condition~(\ref{21}) is marginally satisfied by the Polytropic
gas. Finally, since $P=P(T)$, $(\frac{\partial ^{2}P}{\partial
V^{2}})_{T}$ and $(\frac{\partial ^{3}P}{\partial V^{3}})_{T}$ are
also zero which means that there is no critical point in this
situation \cite{CALEN}. It is useful to note that this conclusion
is the direct result of using Eq.~(\ref{32}) to obtain $b$. By
inserting Eq.~(\ref{39}) into Eqs.~(\ref{EoS})
and~(\ref{thermodynamics}), we get
\begin{equation}\label{530}
\rho
=\frac{(-K)^{-n}}{[1+((\frac{T_{*}}{T})^{\frac{1}{n+1}}(-1)^{\frac{n}{n+1}}-1)^{-1}]^{n}},
\end{equation}
and
\begin{equation}\label{520}
U=\frac{(-K)^{-n}V}{[1+((\frac{T_{*}}{T})^{\frac{1}{n+1}}(-1)^{\frac{n}{n+1}}-1)^{-1}]^{n}},
\end{equation}
as the energy density and total energy of Polytropic gas,
respectively. As we know, the entropy of a thermodynamical system
should be positive \cite{CALEN}. Moreover, Eqs.~(\ref{39})
and~(\ref{530}) can be used to get
\begin{eqnarray}\label{omega}
\omega=-\frac{1}{1+((\frac{T_{*}}{T})^{\frac{1}{n+1}}(-1)^{\frac{n}{n+1}}-1)^{-1}},
\end{eqnarray}
as another relation for the state parameter of Polytropic gas.
This equation indicates that for a Polytropic gas with Polytropic
index satisfying the $(-1)^{\frac{n}{n+1}}=1$ condition and
temperature $T$, which meets the $0 \leq T \leq T_*$ range, the
state parameter encounters the $-1 \leq \omega \leq 0$ range. By
inserting this equation into Eq.~(\ref{35}), one reach
\begin{eqnarray}\label{ent1}
S=\frac{V}{K^nT_*}(-\frac{\omega}{\omega+1})^n,
\end{eqnarray}
for entropy. Hence, since from Eq.~(\ref{press0}) $\omega$ meets
the $-1\leq \omega \leq 0$ condition, the $S>0$ condition is met
if $K^nT_*>0$. Now, using Eq.~(\ref{37}) to get
\begin{equation}\label{38}
C_{V}=n\frac{V}{K^{n}T_{*}}(-1)^{\frac{n}{n+1}}(\frac{T_{*}}{T})^{\frac{1}{n+1}}
[(\frac{T_{*}}{T})^{\frac{1}{n+1}}(-1)^{\frac{n}{n+1}}-1]^{n-1},
\end{equation}
for the heat capacity at constant volume, which can be rewritten
as
\begin{equation}\label{388}
C_{V}=n\frac{S(1+(\frac{V}{\delta})^{\frac{1}{n}})^{\frac{1}{n+1}}}{K\rho^{\frac{1}{n+1}}},
\end{equation}
where we have used Eqs.~(\ref{2444}) and~(\ref{35}) to obtain this
equation. Now, inserting Eq.~(\ref{ent1}) into this equation to
obtain
\begin{equation}\label{388}
C_{V}=n\frac{V(1+(\frac{V}{\delta})^{\frac{1}{n}})^{\frac{1}{n+1}}}{K^{n+1}T_*\rho^{\frac{1}{n+1}}}
(-\frac{\omega}{\omega+1})^n.
\end{equation}
Therefore, the $C_{V}>0$ condition is satisfied if
$\frac{n}{K^{n+1}T_*\rho^{\frac{1}{n+1}}}>0$. By combining this
result with the result obtained from the $S>0$ condition, we get
if the $\frac{n}{K\rho^{\frac{1}{n+1}}}>0$ condition is satisfied,
then the $S>0$ and $C_V>0$ conditions are simultaneously
satisfied.

\subsection*{Some examples}
Consider a Polytropic gas with $n=-\frac{2k}{2k+1}$, where $k$ is
an integer and positive number, which leads to $\rho>0$. From
Eq.~(\ref{19}), it is obvious that this gas satisfies
condition~(\ref{18}). Eq.~(\ref{ent1}) indicates $T_*>0$ to
preserve the $S>0$ condition which yields $b>0$~(\ref{32}). In
addition, from Eq.~(\ref{388}), we see that the $C_V>0$ condition
leads to $K<0$. Bearing Eq.~(\ref{omega}) in mind, since
Eq.~(\ref{press0}) indicates $-1 \leq \omega \leq0$, we conclude
that temperature should meet the $0 \leq T \leq T_*$ condition.
Therefore, this Polytropic gas behaves as a pressureless matter
for $T\rightarrow T_*$ and small volumes. Moreover, its state
parameter and temperature are decreased to $-1$ and $0$,
respectively, by increasing the volume. During this process, the
Polytropic gas satisfies the thermodynamic stability condition.
Loosely speaking, this behavior is similar to the universe
expansion history from the matter dominated era to the current
accelerating phase \cite{roos}. It is useful to note that the
generalized Chaplygin gas has also similar behavior as that of
mentioned here \cite{plb1}.

As another example, focus on a Polytropic gas with positive
density, which yields $P<0$, and $n=-\frac{2k+1}{2k+3}$, where $k$
is again a positive integer number. Since we assumed $\rho>0$,
Eq.~(\ref{8}) implies $K<0$. Eq.~(\ref{19}) shows that
condition~(\ref{18}) is fulfilled. Bearing the $\delta$ definition
in mind, in order to get positive values for $\delta$, $b$ should
meet the $b<0$ condition. Considering Eq.~(\ref{32}), the latter
leads to $T_*<0$. From Eqs.~(\ref{ent1}) and~(\ref{388}) it is
apparent that the $S>0$ and $C_V>0$ conditions are simultaneously
met in this situation, respectively. Bearing Eq.~(\ref{omega}) in
mind, in order to preserve the $-1 \leq \omega \leq0$ condition,
which comes from Eq.~(\ref{press0}), one gets $0 \leq T \leq -T_*$
as a permissible range for temperature. The latter is fully
consistent with physical expectance about positivity of $T$.
Finally, we may conclude that the investigated Polytropic gas may
provide a suitable model for describing the sources of universe
expansion in the matter and current accelerated eras.

Finally, consider a Polytropic gas with $n<-1$ and $K<0$.
According to~(\ref{8}) and~(\ref{11}), it is obvious that $\rho>0$
and $P<0$, respectively. From Eq.~(\ref{19}), it is apparent that
condition~(\ref{18}) is not satisfied. Also, it is easy to show
that the $S>0$ and $C_V>0$ conditions may be met. For example, if
$T_*>0$ and $K^n>0$ then the $S>0$ and $C_V>0$ conditions are
satisfied. Bearing Eqs.~(\ref{press0}) and~(\ref{omega}) in mind,
this gas behaves as a pressureless matter and a fluid with
$\omega\simeq -1$ for small volumes and at the $T\rightarrow T_*$
limit and for large volumes and at the $T\rightarrow 0$ limit,
respectively. It is shown that such a Polytropic gas may lead to a
satisfactory fitting with the observational data \cite{nK}.

\section{The Polytropic gas parameters from thermodynamic arguments}
Consider a Polytropic gas with initial condition $V=V_0$, $P=P_0$,
$\rho=\rho_0$ and $T=T_0$. From Eq.~(\ref{thermodynamics3}) we get
\begin{equation}\label{42}
b=-(\rho _{0}^{-\frac{1}{n}}+K)V_{0}^{-\frac{1}{n}}.
\end{equation}
Now, inserting this result into Eqs.~(\ref{8}) and~(\ref{EoS}) to
obtain
\begin{equation}\label{43}
\rho=\rho_{0}(-K\rho _{0}^{\frac{1}{n}}+(1+K\rho
_{0}^{\frac{1}{n}})(\frac{V}{V_{0}})^{\frac{1}{n}})^{-n},
\end{equation}
and
\begin{equation}\label{44}
P=K \rho ^{1+\frac{1}{n}}=K \rho _{0}^{1+\frac{1}{n}}(-K \rho
_{0}^{\frac{1}{n}}+(1+K \rho
_{0}^{\frac{1}{n}})(\frac{V}{V_{0}})^{\frac{1}{n}})^{-(n+1)},
\end{equation}
for the density and pressure of Polytropic gas, respectively.
Here, we define new parameters $\varepsilon$, $v$, $p$, $\gamma$,
$t$ and $t_*$ as
\begin{eqnarray}\label{45}
\varepsilon&=&\frac{\rho}{\rho_{0}}, \ \ \ v=\frac{V}{V_0},\ \ \
p=\frac{P}{K^{-n}}, \nonumber \\ \gamma&=& -K\rho
_{0}^{\frac{1}{n}},\ \ \ t=\frac{T}{T_{0}},\ \ \
t_{*}=\frac{T_{*}}{T_{0}}.
\end{eqnarray}
It is useful to mention that $\gamma$ is nothing but
$\widetilde{K}$ defined in \cite{nK}. Therefore, Eqs.~(\ref{43})
and~(\ref{39}) can be written as
\begin{equation}\label{47}
\varepsilon =(\gamma +(1-\gamma )v^{\frac{1}{n}})^{-n},
\end{equation}
and
\begin{equation}\label{46}
p=[(-(1+[(\frac{t_{*}}{t})^{\frac{1}{n+1}}(-1)^{\frac{n}{n+1}}-1])^{-1})]^{-(n+1)},
\end{equation}
respectively. Moreover, from Eq.~(\ref{44}) we obtain
\begin{equation}\label{48}
p=\frac{P}{K^{-n}}=K^{n+1}\rho ^{\frac{n+1}{n}}=
\frac{(-1)^{n+1}\gamma
^{n+1}}{(\gamma+(1-\gamma)v^{\frac{1}{n}})^{n+1}}.
\end{equation}
Inserting initial conditions into Eqs.~(\ref{46}) and~(\ref{48}),
and equating them to each other to get
\begin{equation}\label{50}
t_{*}=\frac{(-1)^n}{(1-\gamma)^{n+1}}.
\end{equation}
In fact, this equation helps us to relate $K$, $n$, $T_*$, $T_0$
and $\rho_0$ to each other. Now, consider a Polytropic gas with
$\rho>0$ and $n=-\frac{2k}{2k+1}$, where $k$ is an integer and
positive number. Now, combining Eqs.~(\ref{45}) and~(\ref{50}) to
obtain
\begin{eqnarray}
K=[(\frac{T_0}{T_*})^{\frac{1}{2k+1}}-1]\rho_0^{1+\frac{1}{2k}}.
\end{eqnarray}
Therefore, for $0\leq T_0<T_*$, since $k>0$, this equation
indicates $K<0$ which is in line with the results obtained in the
previous section. Indeed, for every Polytropic index $n$ we get
\begin{eqnarray}
K=\frac{(\frac{T_0}{T_*})^{\frac{1}{n+1}}(-1)^{\frac{n}{n+1}}-1}{\rho_0^{\frac{1}{n}}},
\end{eqnarray}
which indicates that in order to get real values for $K$, whiles
$T_*>0$, $n$ should meet the $(-1)^{\frac{n}{n+1}}=\pm1$
condition. For example, if $0\leq T_0<T_*$ and $n$ meets the
$(-1)^{\frac{n}{n+1}}=1$ condition, we get $K<0$ and thus
$\gamma>0$ meaning that $\widetilde{K}>0$ which is fully
consistent with the results obtained in Ref. \cite{nK}. Loosely
speaking, by applying some initial conditions on the Polytropic
gas we get a relation between the initial conditions and the
Polytropic gas parameters including the Polytropic index ($n$) and
constant $K$.
\section{Summary and concluding remarks}
Here, we have considered an adiabatic Polytropic gas and found out
its energy density and pressure by using pressure definition in
thermodynamics and the equation of state of Polytropic gas,
respectively. Thereinafter, we got the corresponding relations for
the state parameter as well as the total energy of Polytropic gas.
In addition, by investigating the asymptotic behavior of energy
density of Polytropic gas for small and large volumes, and using
the positivity of its density for even Polytropic indexes
condition \cite{cris,ref59,taji}, we got the $\delta>0$ condition
for parameter $\delta$ appeared in relations. Our study shows that
the state parameter of Polytropic gas meets the $-1\leq \omega\leq
0$ range for the finite values of $\delta$. In continue, we
reviewed the thermodynamic stability condition which comes from
the energy surfaces convexity condition \cite{CALEN}. In order to
investigate all of the stability conditions, we used dimensional
analysis to get a primary relation for $b$ and thus temperature.
Due to our special choice for temperature, no critical point
observed for Polytropic gas. We should note that this result may
be changed by choosing different functions for $b$. We also
pointed to the entropy of Polytropic gas, and get to the
$K^nT_*>0$ condition to have positive entropy. Our study shows
that whenever the $\frac{n}{K^{n+1}T_*\rho^{\frac{1}{n+1}}}>0$
condition is satisfied, the $C_V>0$ condition is also met by
Polytropic gas. We have pointed to the behavior of some Polytropic
gases, and found out that a Polytropic gas with
$n=-\frac{2k}{2k+1}$, where $k$ is a positive integer number, and
$\rho>0$ satisfies all of the stability conditions if $T_*>0$ and
$K<0$. Under such criterion, the state parameter and temperature
of Polytropic gas meet the $-1\leq \omega \leq 0$ and $0\leq T
\leq T_*$ ranges, respectively. Polytropic gas with
$n=-\frac{2k+1}{2k+3}$, where $k$ is a positive integer number,
and $K<0$ is also addressed. We found out that the $S>0$ and
$C_V>0$ conditions indicate $T_*<0$. Finally, we saw that its
temperature should meet the $0\leq T \leq -T_*$ range in order to
preserve the $-1\leq \omega \leq 0$ condition obtained from
Eq.~(\ref{press0}). Our study also shows that a Polytropic gas
with $n<-1$ and $K<0$ may satisfy the $S>0$ and $C_V>0$ conditions
whiles $T_*>0$ and $\delta$ is considered as a finite and
well-defined quantity. Such a Polytropic gas has positive density
and behaves as a pressureless matter for small volumes and at the
$T\rightarrow T_*$ limit and a fluid with $\omega \rightarrow -1$
for large volumes and at the $T\rightarrow 0$ limit, whiles
$K^n>0$. It is worthwhile to mention here that a Polytropic gas
with $n<-1$ and $K<0$ has an appropriate fitting with
observational data \cite{nK}. Finally, by imposing some initial
conditions to under investigation Polytropic gas, we could find a
relation between the Polytropic gas parameters, including $n$ and
$K$, and the initial conditions applied on system.
\acknowledgments{The work of H. M. has been supported financially
by Research Institute for Astronomy \& Astrophysics of Maragha
(RIAAM) under research project No. $1/4165-5$.}

\end{document}